# Evanescent Wave Dynamic Light Scattering of Turbid Media

Antonio Giuliani*, Benoit Loppinet@

*Institute for Electronic Structure and Laser,
Foundation for Research and Technology Hellas, Heraklion Greece*

*Present address : University of Groningen, the Netherlands
@ benoit@iesl.forth.gr

## Abstract

Dynamics light scattering (DLS) is a widely used techniques to characterize dynamics in soft phases. Evanescent Wave DLS refers to the case of total internal reflection DLS that probes near interface dynamics. We here investigate the use of EWDLS for turbid sample. Using combination of ray-tracing simulation and experiments, we show that a significant fraction of the detected photons are scattered once and has phase shifts distinct from the multiple scattering fraction. It follows that the measured correlation can be separated into two contributions: a single scattering one arising from the evanescent wave scattering, providing information on motion of the "scatterers" and the associated near wall dynamics and a multiple scattering contribution originating from scattering within the bulk of the sample. In case of turbid enough samples, the latter provides diffusive wave spectroscopy (DWS) -like correlation contribution. The validity of the approach is validated using turbid colloidal dispersion at rest and under shear. At rest we used depolarized scattering to distinguish both contributions. Under shear, the two contributions can easily be distinguished as the near wall dynamics and the bulk one are well separated. Information of both the near wall flow and the bulk flow can be retrieved from a single experiment. The simple structure of the measured correlation is opening the use of EWDLS for a large range of samples.

## Introduction

Dynamic light scattering (DLS) with its many implementations is a pillar of experimental soft condensed matter [1]. It sets the standard for particle sizing. Its extended timescale range has been pivotal in the elucidation of complex dynamics. Scattered intensity collected at a detector is fluctuating in time as a result of the motion of the scatterers. The intensity auto-correlation function (IAF) $\langle I(t)I(t+\tau)\rangle$, or field autocorrelation function (FAF) $\langle E(t)E(t+\tau)\rangle$ where $I(t) = |E|^2(t)$ offer a convenient way to analyze the fluctuation and to obtain information on the motion. In the original form of single scattering DLS, fluctuation at a specific wave vector ***q*** (set be scattering angle) are probed and the FAF directly relates to the intermediate scattering function $S(q,t)$. In the presence of strong multiple scattering, the detected photons can be considered to have diffused through the sample with a transport mean free path $l^*$. The fluctuations and the DLS auto-correlation function then relate to microscopic motion over distance $l^*/N$ where *N* is the average number of scattering events. It is then known as diffusing wave spectroscopy (DWS) [2].

We here treat of DLS ability to address interfacial dynamics. Evanescent wave DLS (EWDLS) designates a special case of DLS where the incident beam undergoes total internal reflection (TIR) at an interface between a substrate (generally glass) and the sample. An exponentially decaying field (evanescent wave) penetrates the sample with penetration depth $dp$. The scattered photons collected at a detector in the far field originate from the evanescent field and the IAF reflects the dynamics in the vicinity of the interface. The typical EWDLS experiment geometry is sketched in Fig. 1A. So far its use has been limited to single scattering [3]–[7]. It imposes limitations in term of samples, and reduces the intensity to rather low values. Here, we explore the use of EW-DLS in the case of turbid samples. We show how both near wall dynamics and bulk dynamics both contribute to the IAF and how the two contributions can be distinguished.

To understand this, a simple symmetry argument can be made, considering the illumination along the surface characteristic of evanescent waves, as sketched in Fig 1B. Assuming that the first scattering event is to take place within d$p$, i.e. within the evanescent field, the scattered photon wave vector can point either towards the glass or towards the sample bulk with equal probability. Under the condition of small illuminated volume ($dp \ll l^*$, beam diameter $\leq l^*$), half of the single scattered photons should be scattered back to the glass and exit the substrate without further scattering (minus reflection losses). The other half should follow multiple scattering paths and eventually escape the sample and part of it will hit the detector. Under the point of entrance detection, a large part (close to half) of the detected photons should therefore reach the detector after a single scattering event, and provide EWDLS like IAF. The other photons will be multiply scattered and provide DWS like IAF (at least if $L \gg l^*$). If the two contribution can be easily separated, both could be analyzed and the near wall dynamics could possibly be retrieved.

In the following we evaluate the validity of the argument using ray tracing simulations and experiments. We show that indeed near wall dynamics information can be retrieved from experimentally measured IAF. A situation where the two contributions should be easy to distinguish is when the dynamics near the wall and in the bulk are different. A simple way to achieve this is through shear flow where velocities are proportional to their distance from the interface. The velocity differences will reflect clearly in the IAF. In the case of homogeneous dynamics like diffusion in a sample at rest, distinguishable phase distributions between the single and the multiple scattering photons will arise from the difference in number of scattering events.

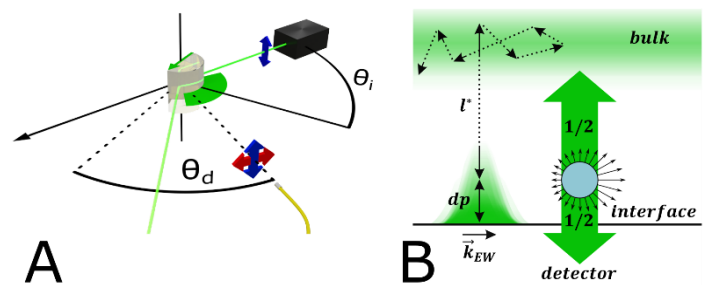

**Fig. 1.** A: Diagram of a typical EWDLS experiment. The incident angle $\theta_i$ (measured from the surface normal) sets the penetration depth $dp$ when $\theta_i > \theta_c$. The detection angle (measured from the surface) $\theta_d$ sets the scattering wave vector. The incoming polarization is V (blue arrow) and a analyzer allows for VV (blue arrow) or VH (red arrow) detection. B: Sketch of scattering distribution from an EW by a turbid medium. Roughly half of the photons scattered once goes outside of the sample volume from the substrate side (and some reach the detector) and half propagates into the turbid medium through higher order scattering. The distance between consecutive scattering event is represented by the dashed arrows with an average length $l^*$ ($\gg dp$).

## Ray tracing simulations

Ray tracing simulations were based on the code provided by [8], [9] for 3D Mueller matrix, Mie scattering, Monte Carlo polarized light transport using quaternions. The code was modified to mimic the EW experimental conditions. Photons were initiated with a wave vector $\vec{k}_{EW}$ along the interface and an exponentially decaying intensity distribution. The characteristics of the optical fiber detection were mimicked. The beam diameter of 0.5mm was used as in experiments, but a wider acceptance angle was used (1° vs 0.13°) to increase the count rate. Photons were collected at a total of 3 discrete angles around the nominal scattering angle (e.g. (60°, 61°, 62°) for 61°). Reflection and transmission Mueller matrices of the medium-glass interface were also considered.

The paths of detected photons were collected. The phase shifts $\Delta\phi_{s,\{D,F\}}(\tau)$ for a photon path $s$ were calculated as $6\sum_i D q_i^2 \tau$ (1a) and $\sum_i v(r_i) q_i \tau$ (1b) for diffusion and flow respectively [10], [11]. $q_i$ is the scattering wave vector of the $i^{th}$ scattering event on the photon path. The particle diffusion coefficient $D$ was kept constant. Simple linear shear flow was considered with $v(r) = \dot{\gamma}z$ with the shear rate $\dot{\gamma}$ constant. No near-wall hydrodynamic effects were considered. FACF of the detected photons were calculated by using the phase argument and a photon weight $W$. $W$ is initially 1 and decreases to mimic absorption, reflection, and transmission losses. The field correlation are computed as

$$g_{1D}(\tau) = \frac{\sum_s W_s \exp(-\Delta\phi_{s,D}(\tau))}{\sum_s W_s}, \quad g_{1F}(\tau) = \frac{\sum_s W_s \exp(-i\Delta\phi_{s,F}(\tau))}{\sum_s W_s} \quad \text{(Eq. 2a ,2b)}$$

for diffusion and flow respectively.

A large enough number of photon was launched ($10^7$) to allow for sufficient scattering events to be recorded for statistics (250k in 22k paths for flow, and 270k in 7k paths for diffusion). Note that the simulations shown use the experimental parameter described below (concentration, refractive indexes, dimensions, and $\theta_i$ and $\theta_d$).

## EWDLS experiments

Two separate experimental setups were used for samples at rest or under shear flow. Light sources were 532 nm lasers. Detection was done from the same side as the illumination with a collimator and optical fiber pointing at the EW, feeding a PMT. A polarizer placed between the sample and the optical fiber was used to select between co-polarized (VV) and cross-polarized (VH) light. Polarization is used to distinguish different paths: Single scattered light keeps the polarization of the incoming beam. Multiple scattered light is depolarized so that VV contains both single scattering and multiple scattering and VH contains only multiple scattering. The IAF was obtained by hardware digital correlators (ALV5000, correlator.com). FAF at rest were obtained from measured IAF by solving (the generalized Siegert relation) $g_2(\tau) = 2|g_1(\tau)|^2(1-B)^2 + 2Re(g_1(\tau))B(1-B)$ where $B = \sqrt{1 - g_2(\tau \to 0)}$. Flow IAF are shown normalized by amplitude at $\tau \to 0$ (this operation preserves the near wall velocity of interest). In this case, comparison with simulations were done by calculating simulated IAF as $g_2(\tau) = (|(1 - I_0)g_1(\tau) + I_0|^2 - I_0^2)/(1 - I_0^2)$ accounting for the static coherent intensity $I_0$ (taken = 0.1) originating from surface defects in real experiments [1], [7].

EWDLS at rest was performed using a two arms single axis goniometer that allowed independent control of incidence angle and of the scattering angle (see Fig. 1A). The sample cell was made of a semi-cylindrical lens and a flow cell with thickness $\sim 250\ \mu m$. PS (n=1.59) spheres of 190 nm in diameter were used as scatterers. A 10%w dispersion in water (n=1.33) was used with $l^*$=20 μm. The critical angle imposed by the ratio of sample and substrate refractive index was at 61.0° from normal incidence. The incidence angle was set at $\theta_i = 66°$ corresponding to a nominal field $dp$ of 210 nm, and the detection angle at $\theta_d = 61°$ from grazing angle.

EWDLS under flow was performed on a rheometer (Malvern Kinexus Pro+) equipped with parallel plates. A comprehensive set-up description can be found in [12]). TIR is realized at the fixed BK7 glass bottom plate, a fused silica 34 mm diameter top plate was used, and the gap was 450 μm. Detection was done at a single scattering angle perpendicular to the interface (90° scattering angle). 5%w dispersion in a water-glycerol mixture (n=1.438) of the same PS probe spheres with $l^*$=120 μm; the critical angle was 71.09°, incidence angle was $\theta_i = 72.8°$, the nominal field $dp$ was 420 nm and the scattering angle was $\theta_d = 90°$.

## Results and Discussion

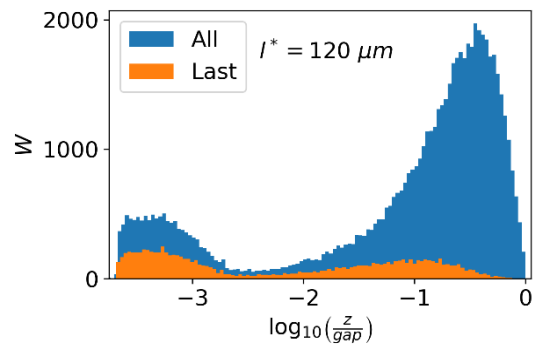

**Fig. 2.** Weighted histogram of z coordinates for the last scattering event before detection and for all the events for $l^* = 120\ \mu m, gap = 1\ mm, dp = 0.5\ \mu m$.

We first consider the spatial distribution of scattering events within the samples as obtained from the ray tracing simulation. On Fig. 2 the distribution of all the scattering events along the $z$ direction (perpendicular to the interface) is plotted in log scale (in blue) together with the 'last' event before detection distribution (in orange). Both distributions are clearly bimodal with a clear separation of the scattering happening within the EW ($< -2.5$ first order scattering) and those from the bulk sample (($> -2.5$ higher order scattering). Intermediate values in $log(z)$ are mostly unpopulated when $dp \ll l^*$ so that the photons probe either exclusively the near wall (single scattering) or the bulk.

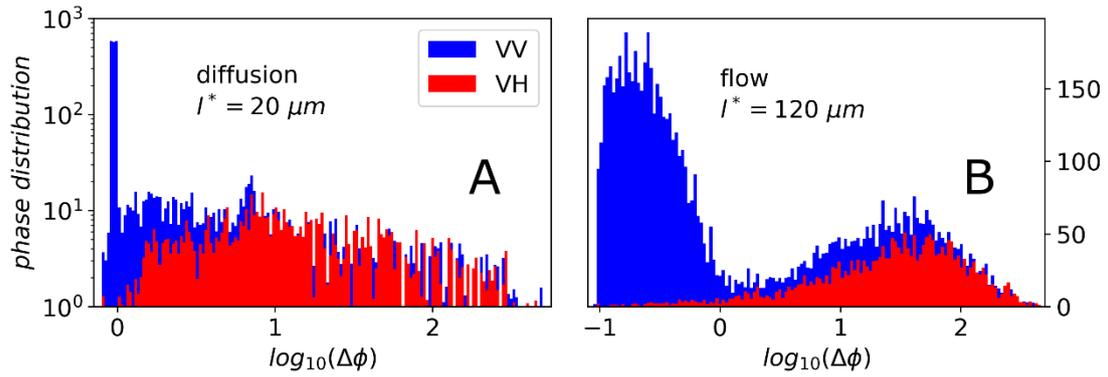

**Fig. 3.** Phase shifts of simulated photons for both polarizations weighted with their respective $W$. A: Diffusion for $l^*$= 20 μm in a 250 $\mu m$ thick cell and $\theta_d = 61°$. B: Flow for $l^*$=120 μm in a 450 μm thick cell and $\theta_d = 90°$.

The distribution of Fig. 2 can be transformed onto phase distribution and onto IAF by accumulating the phase shift of each scattering event along a scattering path (Eq. 1a-1b). Phase shift distributions of the detected photons are shown in Fig. 3 for the case of diffusing particles (A) (using log y-axis) and for the case of flowing particles (B) (with linear y-axis) for both VV and VH scattering. In the diffusion case, the VV distribution of phase shows a strong (Dirac-like) peak two orders of magnitude higher than the baseline shown as seen in Fig. 3A. It corresponds to the single scattering with $Dq^2$ phase shift as would be seen in optically dilute EWDLS. The base signal spreads over two decades and corresponds to the higher order scattering. Expectedly, the single scattering peak is absent from the VH and only a broad distribution of larger phase angles is present. In this case, the dynamics are homogeneous over the sample so the local phase variations are not strong throughout the sample and the broad tail peaks around $\Delta \phi = 10$. The spread of the phase is mostly the result of the different path lengths (number of scattering events) of the detected photons. In the sheared dispersion, the velocity $v$ varies linearly with $z$ according to the shear rate so an event near the wall can be up to three orders of magnitude slower than a bulk event ($dp/gap \sim 10^{-3}$). $dp \ll l^*$ almost guarantees that single scattering events will have a significantly smaller phase shift. The phase distributions for the sheared dispersion (shown for VV and VH in Fig. 3 right) clearly show well separated near wall and the bulk velocimetry modes in VV. The low VV phase shift mode reflects the in penetration depth shear rate as would be measured in single scattering EWDLS [7], [12]. The larger phase shift reflects bulk shear flow measured through a DWS like method (as the final phase reflects the phase addition of the different scattering event). It should reflect the velocity distribution in the bulk samples. The relation to the exact flow profile would depend on the experimental parameters as in the case of DWS under flow [11], [13]. Both diffusion and flow show bimodal distributions. The narrow distribution (with Dirac like peak) observed in the diffusion case is a result of the homogenous dynamics, it is spread over a broader peak in the case of flow due to the velocity gradient. The phase distribution Fig. 3A-3B suggests that a subtraction of the VH contribution in the VV one would lead to the retrieval of low phase shift contribution that essentially relates to the near wall single scattering. This is confirmed as the VV-VH subtraction distribution is found to agree well with single scattering (to better than 90 %) even in the diffusive case. The small difference is expected to be multiple scattering that would affect the low phase shift part of the (VV-VH) distribution.

The phase distribution can be transformed into time domain ACF using Eq. 2a-b . They can be compared to the experimentally measured ACF. The simulated and experimental VV and VH FACF are shown in Fig. 4A and 4B respectively in the case of diffusion. Note that in Fig. 4B the time axis is rescaled by $(q^2 + dp^{-2})$ as the single scattering experiment was carried out at a different $dp$. The $g_{1VH}$ decays faster than $g_{1VV}$ as expected from phase distribution in simulation, is also observed experimentally. The qualitative agreement between simulation and experiments is striking, though the shape of the experimental ACF appears different from the simulated ones. This is probably the result of simplifications done in the simulation (like not including hydrodynamic wall effects).The VV-VH subtraction operation suggested above has been attempted for both experimental and simulated FAF. Results are shown in Fig4 together with single scattering FAF for comparison. The experimental single scattering FAF was measured with optically diluted dispersion of the same probes. The subtraction was weighted by the respective intensities, $g_{1sub} = g_{1VV} - g_{1VH}\left(\frac{A_{VH}I_{VH}}{A_{VV}I_{VV}}\right)$ to account for intensity effects. . For the simulation the total intensity ($I = \sum_s W_s$) is used. As in the case of simulation, the agreement between the subtraction and the single scattering is excellent confirming the validity of the approach.

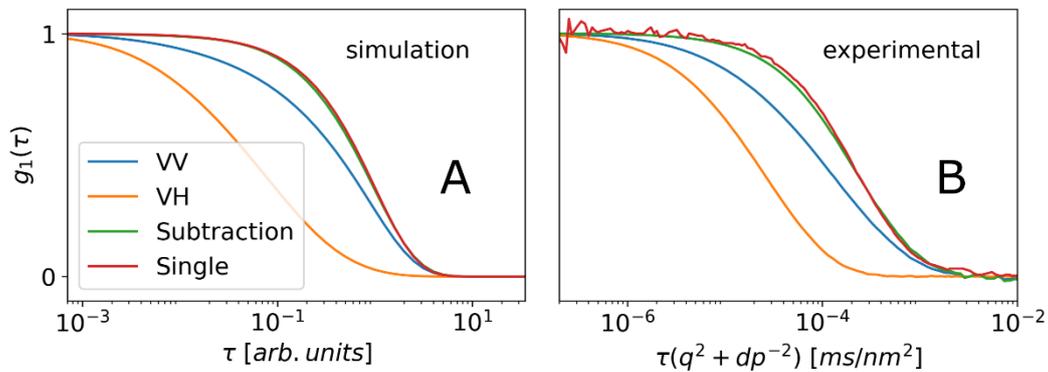

**Fig. 4.** Simulated (A) and experimental (B) FAF for diffusing particles, $l^*$=20: Also shows the single scattering and the estimated single scattering from subtraction VV- VH procedure.

Results of simulation and experiments in the case of applied shear flow are shown in Fig. 5. The VV IAF adopt a clear two step decay (for large enough shear rate) in both simulation (Fig. 5D) and experiments (Fig. 5A and C). The experimental normalized IAF superimpose when the time axis is scaled by the imposed shear rate $\dot{\gamma}$ in Fig. 5C showing that the shear rate it is the relevant time constant for both the slow and fast modes. The slow mode is only present in the polarized VV signals and represents the near wall mostly single scattering contribution. In single scattering regime, the IAF result from the velocity distribution within the $dp$. Two quantities can be extracted, the local near wall shear rate, and the slowest velocity of the scatterers [7], [12]. The functional forms of the VV slow modes observed with turbid dispersions are in essence similar to the ones encountered in single scattering near wall velocimetry, so that the same analysis can be used to retrieve the near wall velocity and shear rate. A minor contamination of low order multiple scattering terms can be found in the faster channels but this has no important effect on determining the near wall velocities. The similarity of near wall velocities measured in a Newtonian optically dilute sample (low probe concentration) and on an equivalent Newtonian turbid sample is apparent in Fig. 5B where both are displayed as a function of shear rates. They both correspond to stick boundary conditions for the flow. We should add that measurement times are significantly reduced and statistics are improved in the turbid case owing to the much higher intensities available in optically dense samples. The fast multiple scattering mode present in both VV and VH polarizations is related to the bulk flow. A proper analysis in terms of DWS of flowing systems will be needed to possibly retrieve precise information on velocity profiles [11], [13], [14]. This would be of great interest but lies outside the scope of this article. We nonetheless noticed that the VV and VH characteristic decay times show specific dependence on applied shear rate $\dot{\gamma}$ (time scale) similar to the DWS under flow. The VH characteristic time was found to scale as $\langle \tau \rangle \propto (gap\, \dot{\gamma})^{-1}$ almost independently of $l^*$. The VV mode was also observed to follow the same trend for large enough $l^*$. It corresponds to the forward scattering DWS case. For shorter $l^*$, $\langle \tau_{VV} \rangle \propto (l^* \dot{\gamma})^{-1}$ as seen in the case backscattering DWS.

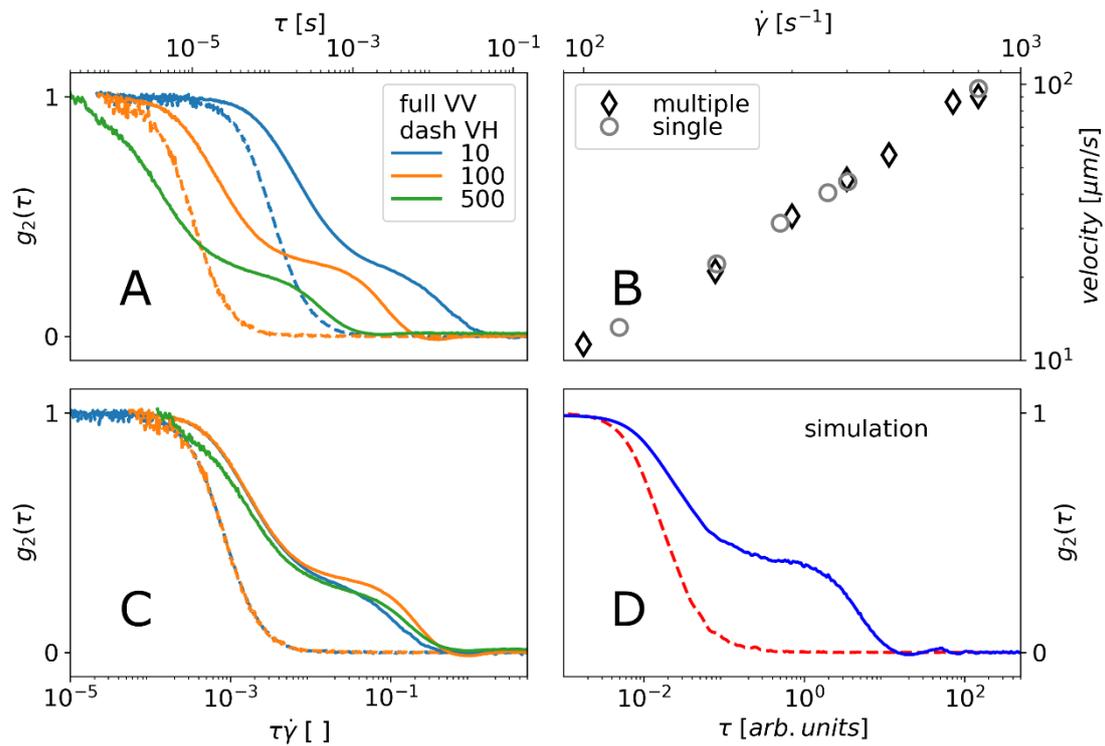

**Fig 5.** A: Flow experimental normalized correlograms measured for different shear rates (10, 100, 500 s$^{-1}$) for VV and VH polarization. C: same data as in A with time axis rescaled by shear rate B: Near wall velocities extracted from the multiple and single scattering samples. D: IAF from ray tracing simulation for VV and VH polarization.

## Conclusion

We have shown using simulation and experiments that EWDLS can be conveniently used in samples with strong scattering. Analyzing the dynamics we proved that the scattering probed through point of entrance detection is composed of two components, a single scattering one and a multiple scattering one. We showed the both contribution can be extracted from the measured I-ACF. Even for the case of diffusion where the dynamics are homogeneous through the sample, the single scattering correlogram can be retrieved by means of a simple subtraction of the co-polarized field I-ACF from the cross-polarized I-ACF. This is valid at least for samples with no inherent depolarization. In the case of shear flow where the dynamics are dependent of the distance from the wall, the two contributions are clearly distinct in the polarized I-ACF. The fast contribution corresponds to the bulk shear flow and the slow contribution corresponds to the near wall flow. Near wall velocities can be directly extracted from the raw I-ACF without further data manipulations using the method developed for single scattering EWDLS [7], [12]. The findings presented here are the consequence of three conditions: symmetric scattering given the incident wave vector parallel to the interface, strong localization of the EW probe near the interface ($dp \ll l^*$), detection at the point of entrance (at the TIR EW spot). A fourth extra condition, variations in local dynamics as in the case of shear flow, makes it easy to separate the two contributions. The multiple scattering contribution provides information on the bulk dynamics as a special case of DWS. It can bring valuable information on the flow profile as has already been shown in DWS under flow [15].

EWDLS in single scattering is limited by the low scattering intensities from the samples. The large scattering intensities present in turbid samples considerably simplifies the use of EWDLS. Application of EWDLS on turbid samples may open new possibilities of probing interfacial behavior of various turbid samples like emulsion or concentrated colloidal solutions. The addition of scattering probes can turn transparent solutions into turbid ones, as done in DWS microrheology [16]. In this case EWDLS might be able to retrieve to near wall and bulk motion of the probes.

*Acknowledgement*
We acknowledge EU Horizon 2020 funding (EUSMI and SOMATAI) and ESPA 2014-2020 funding (AENO)

*References*
[1] R. Pecora, *Dynamic Light Scattering: Applications of Photon Correlation Spectroscopy*. Plenum Press, New York and London, 1985.

[2] D. Weitz and D. Pine, "Diffusing-wave spectroscopy," in *Dynamic Light Scattering*, W. Brown, Ed. Oxford University Press, 1993, pp. 652–720.

[3] G. E. Yakubov, B. Loppinet, H. Zhang, J. Rühe, R. Sigel, and G. Fytas, "Collective Dynamics of an End-Grafted Polymer Brush in Solvents of Varying Quality," *Phys. Rev. Lett.*, vol. 92, no. 11, p. 115501, Mar. 2004.

[4] E. Filippidi, V. Michailidou, B. Loppinet, J. Rühe, and G. Fytas, "Brownian diffusion close to a polymer brush," *Langmuir*, vol. 23, no. 9, pp. 5139–5142, 2007.

[5] R. Sigel, "Light scattering near and from interfaces using evanescent wave and ellipsometric light scattering," *Curr. Opin. Colloid Interface Sci.*, vol. 14, no. 6, pp. 426–437, Dec. 2009.

[6] B. Cichocki, E. Wajnryb, J. Bławzdziewicz, J. K. G. Dhont, and P. R. Lang, "The intensity correlation function in evanescent wave scattering," *J. Chem. Phys.*, vol. 132, no. 7, p. 074704, Feb. 2010.

[7] B. Loppinet, J. K. G. Dhont, and P. Lang, "Near-field laser Doppler velocimetry measures near-wall velocities," *Eur. Physcal J. E*, vol. 35, no. 7, Jul. 2012.

[8] J. C. Ramella-Roman, S. A. Prahl, and S. L. Jacques, "Three Monte Carlo programs of polarized light transport into scattering media: part I," *Opt. Express*, vol. 13, no. 12, p. 4420, Jun. 2005.

[9] J. C. Ramella-Roman, S. A. Prahl, and S. L. Jacques, "Three Monte Carlo programs of polarized light transport into scattering media: part II," *Opt. Express*, vol. 13, 2005.

[10]     D. A. Weitz, J. X. Zhu, D. J. Durian, H. Gang, and D. J. Pine, "Diffusing-wave spectroscopy: The technique and some applications," *Phys. Scr.*, vol. T49B, pp. 610–621, 1993.

[11]     D. Bicout, E. Akkermans, and R. Maynard, "Dynamical correlations for multiple light scattering in laminar flow," *J. Phys. I*, vol. 1, no. 4, pp. 471–491, Apr. 1991.

[12]     A. Giuliani, R. McKenzie, and B. Loppinet, "Near wall velocimetry on a rheometer," *J. Rheol. (N. Y. N. Y).*, vol. 63, no. 1, pp. 93–104, Jan. 2019.

[13]     X.-L. Wu, D. J. Pine, P. M. Chaikin, J. S. Huang, and D. A. Weitz, "Diffusing-wave spectroscopy in a shear flow," *J. Opt. Soc. Am. B*, vol. 7, no. 1, p. 15, Jan. 1990.

[14]     D. Bicout and R. Maynard, "Diffusing wave spectroscopy in inhomogeneous flows," *Phys. A Stat. Mech. its Appl.*, vol. 199, no. 3–4, pp. 387–411, Nov. 1993.

[15]     A. Raudsepp, P. Callaghan, and Y. Hemar, "A study of the nonlinear rheology of complex fluids using diffusing wave spectroscopy," *J. Rheol. (N. Y. N. Y).*, vol. 52, no. 5, pp. 1113–1129, Sep. 2008.

[16]     T. G. Mason, H. Gang, and D. A. Weitz, "Diffusing-wave-spectroscopy measurements of viscoelasticity of complex fluids," *J. Opt. Soc. Am. A*, 1997.